\documentclass[longauth]{aa}
\usepackage{natbib}
\usepackage{longtable,lscape}
\usepackage{amsmath}
\bibpunct{(}{)}{;}{a}{}{,}
\usepackage{graphicx}

\usepackage[colorlinks=true,citecolor=blue]{hyperref}
\usepackage{color}
\usepackage{xspace}
\usepackage{dcolumn}
\usepackage{longtable}
\usepackage{lscape}
\usepackage{wasysym}
\usepackage{rotating}
\usepackage{afterpage}
\usepackage{mathtools}

\newcommand{\MJup}{M$_{\mathrm{Jup}}$\xspace}

\newcommand{\MSun}{M$_{\odot}$\xspace}

\newcommand{\mic}{$\mu$m\xspace}
\newcommand{\as}{\hbox{$^{\prime\prime}$}\xspace}

\usepackage[varg]{txfonts}
\begin{document}

\title{Constraining masses and separations of unseen companions to
  five accelerating nearby stars\thanks{Based on observation
    made with European Southern Observatory (ESO) telescopes at Paranal
    Observatory in Chile, under programs ID 095.C-0298(A), 095.C-0298(B),
    096.C-0241(A), 096.C-0241(G), 097.C-0865(A), 097.C-0865(D), 1100.C-0481(F),
    1100.C-0481(N), 1100.C-0481(P)}} 
\subtitle{}

\author{D. Mesa\inst{1}, M. Bonavita\inst{1,2}, S. Benatti\inst{3},
  R. Gratton\inst{1}, S. Marino\inst{4,5}, P. Kervella\inst{6},
  V. D'Orazi\inst{1}, S.Desidera\inst{1}, T. Henning\inst{7},
  M. Janson\inst{8}, M. Langlois\inst{9,10}, E. Rickman\inst{11,12},
  A. Vigan\inst{9}, A. Zurlo\inst{13,14,9}, J.-L. Baudino\inst{6},
  B. Biller\inst{7,15,16}, A. Boccaletti\inst{6}, M. Bonnefoy\inst{17},
  W. Brandner\inst{7}, E. Buenzli\inst{7}, F. Cantalloube\inst{9},
  D. Fantinel\inst{1}, C. Fontanive\inst{18,1}, R. Galicher\inst{6},
  C. Ginski\inst{19}, J. Girard\inst{20,17}, J. Hagelberg\inst{21},
  T. Kopytova\inst{7}, C. Lazzoni\inst{1}, H. Le Coroller\inst{9},
  R. Ligi\inst{22}, M. Llored\inst{9}, A.-L. Maire\inst{23,7},
  D. Mouillet\inst{17}, C. Perrot\inst{6}, S. Rochat\inst{17},
  C. Romero\inst{17,24}, D. Rouan\inst{6}, M. Samland\inst{7,8},
  T.O.B. Schmidt\inst{6,25}, E. Sissa\inst{1}, F. Wildi\inst{11}}

\institute{\inst{1}INAF-Osservatorio Astronomico di Padova, Vicolo dell'Osservatorio 5, Padova, Italy, 35122-I \\
  \inst{2}School of Physical Sciences, Faculty of Science, Technology, Engineering and Mathematics, The Open University, Walton Hall, Milton Keynes, MK7 6AA \\
  \inst{3}INAF-Osservatorio Astronomico di Palermo, Piazza del Parlamento, 1, I-90134, Palermo, Italy \\
  \inst{4}Jesus College, University of Cambridge, Jesus Lane, Cambridge, CB5 8BL, UK \\
  \inst{5}Institute of Astronomy, University of Cambridge, Madingley Road, Cambridge CB3 OHA, UK \\
  \inst{6}LESIA, Observatoire de Paris, Universit\'{e} PSL, CNRS, Sorbonne Universit\'{e}, Universit\'{e} de Paris, 5 place Jules Janssen, F-92195 Meudon, France \\
  \inst{7}Max-Planck-Institut f\"ur Astronomy, K\"onigstuhl 17, 69117 Heidelberg, Germany \\
  \inst{8}Department of Astronomy, Stockholm University, AlbaNova, University Center, 109 91 Stockholm, Sweden \\
  \inst{9}Aix Marseille Universit\'{e}, CNRS, LAM (Laboratoire d'Astrophysique de Marseille) UMR 7326, 13388 Marseille, France \\
  \inst{10}CRAL, UMR 5574, CNRS, Universit\'{e} de Lyon, Ecole Normale Superieure de Lyon, 46 All\'{e}e d'Italie, 69364 Lyon Cedex 07, France \\
  \inst{11}Observatoire Astronomique de l'Universit\'{e} de Geneve, 51 Ch. des Maillettes, 1290 Versoix, Switzerland \\
  \inst{12}Space Telescope Science Institute, 3700 San Martin Drive, Baltimore, MD 21218, USA \\
  \inst{13}Nucleo de Astronomia, Facultad de Ingenieria y Ciencias, Universidad Diego Portales, Av. Ejercito 441, Santiago, Chile \\
  \inst{14}Escuela de Ingenieria Industrial, Facultad de Ingenieria y Ciencias, Universidad Diego Portales, Av. Ejercito 441, Santiago, Chile \\
  \inst{15}SUPA, Institute for Astronomy, University of Edinburgh, Blackford Hill, Edinburgh EH9 3HJ, UK \\
  \inst{16}Centre for Exoplanet Science, University of Edinburgh, Edinburgh, UK \\
  \inst{17}Univ. Grenoble Alpes, CNRS, IPAG, F-38000 Grenoble, France \\
  \inst{18}5Center for Space and Habitability, University of Bern, CH-3012 Bern, Switzerland \\
  \inst{19}Anton Pannekoek Institute for Astronomy, University of Amsterdam, Science Park 904, 1098 XH Amsterdam, The Netherlands \\
  \inst{20}Space Telescope Science Institute, Baltimore 21218, MD, USA \\
  \inst{21}Geneva Observatory, University of Geneva, Chemin des Mailettes 51, 1290 Versoix, Switzerland \\
  \inst{22}University of Cote d’Azur, Cote d’Azur Observatory, CNRS, Lagrange Laboratory, France \\
  \inst{23}STAR Institute, Universit\'{e} de Liege, Allee du Six Aout 19c, 4000 Liege, Belgium \\
  \inst{24}ESO Vitacura, Alonso de Cordova 3107, Vitacura, Casilla 19001, Santiago de Chile, Chile \\
  \inst{25}Hamburger Sternwarte, Gojenbergsweg 112, 21029 Hamburg, Germany \\
}

   \date{Received  / accepted }

\abstract
     {}
     {This work aims at constraining the masses and separations of potential substellar companions to five accelerating stars (HIP\,1481, HIP\,88399, HIP\,96334, HIP\,30314 and HIP\,116063) using multiple data sets acquired with different techniques.}
     {Our targets were originally observed as part of the SPHERE/SHINE survey, and radial velocity (RV) archive data were also available for four of the five objects. 
     No companions were originally detected in any of these data sets, but the presence of significant proper motion anomalies (PMa) for all the stars strongly suggested the presence of a companion.
     Combining the information from the PMa with the limits derived from the RV and SPHERE data, we were able to put constraints on the characteristics of the unseen companions.}
    {Our analysis led to relatively strong constraints for both HIP\,1481 and HIP\,88399, narrowing down the companion masses to 2-5~\MJup and 3-5~\MJup and separations within 2-15~au and 3-9~au, respectively. 
    Because of the large age uncertainties for HIP\,96334, the poor observing conditions for the SPHERE epochs of HIP\,30314 and the lack of RV data for HIP\,116063, the results for these targets were not as well defined, but we were still able to constrain the properties of the putative companions within a reasonable confidence level.}
{For all five targets, our analysis has revealed that the companions responsible for the PMa signal would be well within reach for future instruments planned for the ELT (e.g., MICADO), which would easily achieve the required contrast and angular resolution. Our results therefore represent yet another confirmation of the power of multi-technique approaches for both the discovery and characterisation of planetary systems. }


\titlerunning{proper motion}
\authorrunning{Mesa et al.}
   \maketitle
%

\section{Introduction}
\label{intro}

In the last decade the direct imaging (DI) technique has allowed for the detection
of a growing number of planetary mass objects orbiting nearby young stars, such as 51\,Eri\,b \citep{2015Sci...350...64M}, HIP\,65426\,b
\citep{2017A&A...605L...9C}, PDS\,70\,b \citep{2018A&A...617A..44K}, and
PDS\,70\,c \citep{2019NatAs...3..749H}. 
This was made possible in particular thanks to a new generation of high-contrast
imagers mainly devoted to this aim, like the Gemini Planet Imager
\citep[GPI;][]{2014PNAS..11112661M}, VLT/SPHERE \citep{2019A&A...631A.155B}, and
CHARIS \citep{2015SPIE.9605E..1CG}. \par
However, even with such sophisticated instruments, direct detection is limited to giant gaseous companions at large separation (more than 10~au) from the host star. 
Such limitation mainly arises from the challenge of resolving extremely faint sources (with contrast of the order of $10^{-6}$) at relatively close angular separations (a physical separation of 10 au corresponds to ~0.2\as for a star at a distance of 50pc)\par
In addition, recent studies have pointed out the  relative paucity of giant planets (masses
larger than 1~\MJup) at large separations \citep{2019AJ....158...13N,2021A&A...651A..72V} which, combined with the increasing
planetary occurrence rate between 0.1 and 1~au obtained through radial
velocity (RV) surveys \citep[see e.g.,][]{2021ApJS..255...14F} implies that the
expected peak for the distribution of giant planets should reside between 1
and 10~au \citep[e.g.,][]{2018A&A...612L...3M}.\par
While the bulk of the giant planet population therefore appears to be out of reach from current direct imaging surveys, it will likely be the main focus for the future instrumentation of the under construction extremely large telescopes \citep[ELT; see e.g.,][]{2018arXiv180401371P}. There are, however, ways to push the discovery space of the current facilities towards the peak of the giant planet population, and  enhance our comprehension of the formation process of such objects. 
Concentrating on the nearest stars is an obvious solution, but it is also possible to carry on focused programs concentrating the efforts on stars that have higher probability to host detectable companions. 
So far the most successful selection methods for such programs have been those based on proper motion anomalies \citep[PMa or accelerations, see e.g.][]{2018ApJS..239...31B, 2019A&A...623A..72K} defined as the difference between the short-term and the long-term proper motion measured for a star. While these trends have been in the past only used to select targets for binary stars searches, the precision on the proper motion measurements achieved by Gaia \citep{gaia2016} now allows to push towards signals pointing to much smaller companions, down to the sub-stellar and even planetary regime. \par
Several groups \citep[see e.g.][]{2019A&A...623A..72K, 2022A&A...657A...7K, 2018ApJS..239...31B,2021ApJS..254...42B} have recently shared PMa catalogues, including the Hipparcos and $Gaia$ DR2/eDR3 proper motions, as well as a $Gaia$-Hipparcos scaled positional differences, placing all proper motions at the epochs of {\it Gaia} DR2 and eDR3, respectively, with re-calibrated uncertainties for hundred of thousand of stars.
For our work, we choose to use the catalogue from \citet{2022A&A...657A...7K} and, following their approach, we define as accelerating those stars with a PMa  signal-to noise ratio (SNR) larger than 3.
As shown by several recent results, accelerating stars have high probability of hosting companions ranging from low mass stars to sub-stellar companions, including objects close to planetary mass \citep[see e.g.,][]{Bonavita2022, 2020ApJ...904L..25C,2021AJ....162...44S, 2021AJ....162..251C}. 
Even in case of non-detection, the combination of DI data with the information on the PMa can be used to put constraints on the nature of the possible unseen companions causing the acceleration. 
In addition, the availability of radial velocity data provides additional information on closer companions.   
For example, this approach was used by \citet{2021MNRAS.503.1276M} to study  the disk hosting star HD\,107146, leading to the conclusion that the measured PMa should be due to the presence
of a companion with a mass of 2-5~\MJup at a separation of 2-7~au.
Results like this one show how such an analysis is therefore crucial to understand the structure of
planetary systems around these stars and can also be used to define 
a sample of optimal targets for future observations with ELT instruments. \par

In this paper we present the results of our analysis for five stars with strong accelerations pointing towards the presence of a low mass companion and for which DI observations were available from SHINE \citep{2017sf2a.conf..331C,2021A&A...651A..70D,
  2021A&A...651A..71L,2021A&A...651A..72V}. While no companion was retrieved
from the SHINE observations, they have been used to constrain their presence at
separation between 2 and 10~au also using mass limits from archive RV data. A detailed description of each of our targets is provided in Section~\ref{s:sample}, while Section~\ref{s:obs} presents the various data sets used for our analysis. 
Our results are presented in Section~\ref{s:res}  and discussed in Section~\ref{s:dis}. Finally, Section~\ref{s:conclusion} provides some concluding remarks. 


\section{Sample selection}
\label{s:sample}

\begin{table*}
\renewcommand{\arraystretch}{1.4}
  \caption{Main Characteristics of the target stars. Individual references for the stellar ages (expressed as $Age^{max}_{min}$) and masses are provided in Section\ref{s:sample}. Parallaxes and Proper Motion values are from eDR3, from which we also show the RUWE. The values of the proper motion anomaly (PMa), PMa SNR and tangential velocity ($\vec{\Delta v_\mathrm{tan}}$) with the corresponding position angle are from \cite{2022A&A...657A...7K}. \label{t:PMa_values}}
\centering 
\resizebox{\linewidth}{!}{  \begin{tabular}{l|ccccc|cc|cc|cc|cc}
    \hline \hline \noalign{\smallskip}
ID & Age$^{max}_{min}$ & Mass &  SpType & Kmag & Parallax & \multicolumn{2}{c|}{Proper Motion} & \multicolumn{2}{c|}{PMa (HIP-eDR3)} &   \multicolumn{2}{c|}{$\vec{\Delta v_\mathrm{tan}}$ (HIP-eDR3)} & RUWE & SNR$_{PMa}$ \\
& (Myrs) & ($M_{\odot}$) & & & (mas) & (RA: mas/yr) & (Dec: mas/yr) & (RA: mas) & (Dec: mas)  & (m/s) & (PA: deg) & &\\
\hline
HIP 1481 & 45$^{50} _{35}$ & 1.16 & F8V & 6.15 & 23.36 $\pm$ 0.02 & 90.05 $\pm$ 0.02 & -59.21 $\pm$ 0.02 & -0.10 $\pm$ 0.02 & 0.04 $\pm$ 0.02 & 22.13 $\pm$ 6.40 & 290.07 $\pm$ 12.07 & 0.994 & 3.46 \\
HIP 30314 & 149$^{180} _{100}$ & 1.11 & G1V & 5.05 & 41.89 $\pm$ 0.01 & -11.43 $\pm$ 0.02 & 64.68 $\pm$ 0.02 & -0.04 $\pm$ 0.02 & 0.10 $\pm$ 0.02 & 12.52 $\pm$ 3.51 & 336.98 $\pm$ 11.05 & 0.874 & 3.57 \\
HIP 88399 & 24$^{29} _{19}$ & 1.29 & F6V & 5.91 & 20.29 $\pm$ 0.02 & 2.33 $\pm$ 0.02 & -86.23 $\pm$ 0.02 & 0.11 $\pm$ 0.03 & -0.09 $\pm$ 0.02 & 32.66 $\pm$ 8.21 & 130.94 $\pm$ 10.17 & 1.093 & 3.98 \\
HIP 96334 & 150$^{220} _{70}$ & 1.00 & G3V & 6.3 & 26.20 $\pm$ 0.02 & -4.04 $\pm$ 0.01 & -176.05 $\pm$ 0.02 & 0.04 $\pm$ 0.02 & 0.13 $\pm$ 0.02 & 25.13 $\pm$ 5.43 & 14.63 $\pm$ 7.53 & 0.995 & 4.63 \\
HIP 116063 & 300$^{500} _{200}$ & 0.80 & G1V & 5.65 & 33.05 $\pm$ 0.02 & 183.28 $\pm$ 0.02 & -122.68 $\pm$ 0.02 & -0.07 $\pm$ 0.03 & 0.09 $\pm$ 0.03 & 16.54 $\pm$ 5.45 & 324.67 $\pm$ 12.73 & 1.107 & 3.04 \\

\hline\hline
\end{tabular}}
\end{table*}

The starting point for the target selection was the catalogue by \citet{2022A&A...657A...7K}, which lists the value of the proper motion anomaly (PMa), defined as the difference in proper motion between Hipparcos and Gaia eDR3 \citep{2021A&A...649A...1G}, for $\sim$11000 Hipparcos stars. 
From a initial list of targets with PMa with SNR higher than three, we selected the objects with renormalised unit weight error \citep[RUWE; for a more detailed definition and description of its use see][]{2021A&A...649A...2L} higher than 1.4, to avoid the effects of possible degradation of the astrometric parameters. Moreover, a large value of the RUWE parameter is often caused by stellar binarity as discussed in \citet{2022A&A...657A...7K}. 
We then restricted our sample to stars within 50 parsecs, in order to make sure possible companions responsible for the PMa would be accessible with direct imaging. 
The final list of 498 targets was then cross-correlated with the list of targets observed during the
SHINE survey \citep{2021A&A...651A..70D, 2021A&A...651A..71L}. 
After the exclusion of objects for which the presence of a bound stellar companion was known and able to explain the PMa signal, we were left with five targets: HIP\,1481, HIP\,88399, HIP\,96334, HIP\,30314 and HIP\,116063. 
Their main characteristics are summarised in Table~\ref{t:PMa_values} and described in detail in the following Sections. 

\subsection{HIP\,1481}
\label{s:hip1481}

HIP\,1481 (HD\,1466) is an F8 star \citep{2006A&A...460..695T} with a mass of
1.16~\MSun \citep{2021A&A...651A..70D}, located at a distance
of 42.82$\pm$0.03~pc from the Sun \citep{2021A&A...649A...1G}. It has been
recognised as member of the Tucana-Horologium moving group
\citep{2015MNRAS.454..593B}. From this membership \citet{2021A&A...651A..70D}
deduced an age of $45_{-10}^{+5}$~Myr. 
\par
The presence of a debris disk around HIP\,1481 was inferred by
\citet{2014ApJS..211...25C} using Spitzer observations. They modelled the
SED with a double black body with temperatures of 97 and 374~K hinting for a
double belt structure for the disk. Using these results
\citet{2018A&A...611A..43L} found for the inner belt a radius of 0.7~au and for
the outer belt a radius of $\sim$52~au with a gap between the two belts
of around 40~au. Using dynamical models they then concluded that to explain
this disk configuration the presence of at least a planet less massive 
than 3~\MJup and with high eccentricity was required. \par
The PMa retreived from \citet{2022A&A...657A...7K} has a SNR of 3.46, 
making the presence of a companion very likely. They
estimated a mass of 3.20~\MJup for a companion on a 3~au orbit and of
2.55~\MJup for a companion on a 10~au orbit.

\subsection{HIP\,30314}
\label{s:hip54155}

HIP\,30314 (HD\,45270) is a G1 star \citep{2006A&A...460..695T} with a mass of
1.11~\MSun \citep{2021A&A...651A..70D}, located at a distance of
23.87$\pm$0.01~pc. It is part of the AB\,Dor association
\citep{2011ApJ...732...61Z,2018ApJ...856...23G} and it has an estimated age
of $149_{-49}^{+31}$~Myr \citep{2021A&A...651A..70D}.
The SNR of the PMa for HIP\,30314 is 3.57 and \cite{2022A&A...657A...7K}
estimated a mass of 1.66~\MJup for an object orbiting at 3~au from the star and
of 1.40~\MJup for a companion orbiting at 10~au from the star.

\subsection{HIP\,88399}
\label{s:hip88399}

HIP\,88399 (HD\,164249) is a F6 star \citep{2006A&A...460..695T} with a mass of
1.29~\MSun \citep{2021A&A...645A..30Z}, located at a distance of
49.30$\pm$0.06~pc from the Sun \citep{2021A&A...649A...1G}. It is part
of the $\beta$\,Pictoris moving group \citep[e.g.,][]{2017A&A...607A...3M} and
has an estimated age of 24$\pm$5~Myr \citep{2021A&A...651A..70D}. 
HIP\,88399 has a known companion, HD\,164249\,B, an M2 star with an estimated mass of
0.54~\MSun \citep{2021A&A...645A..30Z} and a separation of $\sim$6.5\as
corresponding to $\sim$323~au at the distance of the system
\citep{2021MNRAS.502.5390P}. IR excess was detected using both
WISE \citep{2010AJ....140.1868W}, Spitzer \citep{2014ApJS..211...25C} and
Herschel \citep{2013A&A...555A..11E} data, hinting toward the presence of
debris disk. The disk was finally resolved through ALMA observations
by \citet{2021MNRAS.502.5390P} that estimated for the disk a radius of 63~au
and an inclination lower than $49^{\circ}$. \par
HIP\,88399 has a PMa SNR of 3.98 providing a strong indication of the presence
of a substellar object at short separation from the star or of a larger mass
companion at larger separation.

\subsection{HIP\,96334}
\label{s:hip96334}

HIP\,96334 (HD\,183414) is a G3 star \citep{2006A&A...460..695T} with a mass of
1.00~\MSun \citep{2017A&A...603A...3V} at a distance
of 38.16$\pm$0.03~pc from the Sun \citep{2021A&A...649A...1G}. It is not
part of any known young moving group. Its age has been evaluated of
$150_{-80}^{+70}$~Myr \citep{2017A&A...603A...3V}.
This star was a target both for direct imaging
\citep[see e.g.,][]{2010A&A...509A..52C,2019AJ....158...13N} and RV
\citep[see e.g.,][]{2020A&A...633A..44G} with no detection of low mass
stellar or substellar companion. \par
\citet{2022A&A...657A...7K} lists a PMa SNR of 4.63 that
makes the presence of a companion very probable. 

\subsection{HIP\,116063}
\label{s:hip116063}

HIP\,116063 (HD\,221231) is a G1 star \citep{2006A&A...460..695T} with a mass
of 0.8~\MSun \citep{2017A&A...603A...3V} located at a
distance of 30.25$\pm$0.02~pc from the Sun \citep{2021A&A...649A...1G}. Its age
is $300_{-100}^{+200}$~Myr \citep{2015A&A...573A.126D}. A wide companion for
this star is known \citep[TYC-9339-2158-1;][]{2015A&A...573A.126D} but
its large separation (36.3\as corresponding to more than 1100~au) and 
mass (0.80~\MSun) are not compatible with the detected PMa. The value of the PMa SNR for HIP\,116063 is 3.04, just above the threshold hinting toward the presence of a possible companion.

\section{Observations and data reduction}
\label{s:obs}


\begin{table*}[!htp]
  \caption{List and main characteristics of the SPHERE observations 
     used for this work. }\label{t:obs}
\centering
\resizebox{\linewidth}{!}{
\begin{tabular}{cccccccccc}
\hline\hline
Target & Date  &  Obs. mode & Coronograph & DIMM seeing & $\tau_0$ & wind speed & Field rotation & DIT & Total exp.\\
\hline
HIP\,1481 & 2015-10-26  & IRDIFS      & N\_ALC\_YJH\_S & 1.08\as & 1.4 ms & 1.77 m/s &$25.1^{\circ}$ &  64 s &  4096 s \\
HIP\,1481 & 2016-09-18  & IRDIFS      & N\_ALC\_YJH\_S & 0.80\as & 5.0 ms & 9.95 m/s &$25.1^{\circ}$ &  64 s &  5120 s \\
HIP\,30314 & 2016-01-16 & IRDIFS      & N\_ALC\_YJH\_S & 1.73\as & 1.5 ms & 10.93 m/s &$27.2^{\circ}$ & 64 s & 4096 s \\
HIP\,88399 & 2015-05-10  & IRDIFS      & N\_ALC\_YJH\_S & 1.86\as & 1.2 ms & 2.63 m/s &$34.5^{\circ}$ &  64 s &  3584 s \\
HIP\,88399 & 2015-06-01  & IRDIFS      & N\_ALC\_YJH\_S & 1.25\as & 1.1 ms & 2.48 m/s &$34.1^{\circ}$ &  64 s &  4096 s \\
HIP\,88399 & 2016-04-17  & IRDIFS      & N\_ALC\_YJH\_S & 1.58\as & 1.5 ms & 14.73 m/s &$37.2^{\circ}$ &  64 s &  3776 s \\
HIP\,88399 & 2018-04-11  & IRDIFS      & N\_ALC\_YJH\_S & 0.52\as & 5.6 ms & 7.45 m/s &$32.0^{\circ}$ &  96 s &  3840 s \\
HIP\,88399 & 2019-09-07  & IRDIFS      & N\_ALC\_YJH\_S & 1.14\as & 1.9 ms & 12.75 m/s &$40.7^{\circ}$ &  96 s &  4896 s \\
HIP\,96334 & 2019-05-19  & IRDIFS      & N\_ALC\_YJH\_S & 0.47\as & 4.6 ms & 8.73 m/s &$32.8^{\circ}$ &  96 s &  6144 s \\
HIP\,116063 & 2019-09-07  & IRDIFS      & N\_ALC\_YJH\_S & 2.06\as & 1.5 ms & 18.23 m/s &$31.6^{\circ}$ &  64 s &  4032 s \\
\hline
\end{tabular}}
\end{table*}

\subsection{Direct imaging data}
\label{s:didata}
All our objects were observed as part of the SPHERE/SHINE program. 
Table~\ref{t:obs} shows the main characteristics of the observations, all performed using the IRDIFS SPHERE observing mode, that uses both IFS
\citep{2008SPIE.7014E..3EC} operating in Y and J spectral bands between
0.95 and 1.35~\mic on a $1.7\as\times1.7\as$ field of view (FOV) and IRDIS
\citep{2008SPIE.7014E..3LD} covering in the H spectral band using the H23
filter pair \citep[wavelength H2=1.593~\mic;
  wavelength H3=1.667~\mic;][]{2010MNRAS.407...71V} on a circular FOV of
$\sim$5\as. 
All the observations were performed exploiting the SPHERE adaptive optics system SAXO
\citep{2006OExpr..14.7515F}. \par
At all epochs we acquired frames with satellite spots symmetrical with
respect to the central star just before and just after the science sequence, used to precisely define the position of the star behind the
coronagraph \citep{2013aoel.confE..63L}. To obtain photometric calibration we also observed each star without the coronagraph, using an appropriate neutral density filter to avoid the saturation of the detectors  \par
As detailed in Table~\ref{t:obs}, multiple epochs were available for HIP\,1481 and HIP\,88399. Unfortunately, only one the five data-sets available for HIP\,88399 was taken in good enough weather conditions (2018-04-11). The same is true for HIP\,1481, for which the conditions were significantly better in second night (2016-09-18). 
Therefore we will only use these data sets for our analysis of these two objects.
\par
All data were reduced using the SPHERE data center
\citep{2017sf2a.conf..347D}. The first step was to apply the appropriate
calibrations following the data reduction and handling pipeline
\citep[DRH;][]{2008ASPC..394..581P}. For IRDIS the required
calibrations include the creation of the master dark and of the master flat-field
frames and the definition of the star center. For IFS it was also necessary to define the position of of each spectra on the detector as well as the wavelength calibration and the application of the instrumental flat that takes into account the different response of each lenslet of the IFS array.
On the reduced data we then applied speckle subtraction using both angular
differential imaging \citep[ADI;][]{2006ApJ...641..556M} and spectral
differential imaging \citep[SDI;][]{1999PASP..111..587R}. These methods were implemented using both
the principal components analysis \citep[PCA;][]{2012ApJ...755L..28S} and the
TLOCI \citep{2014IAUS..299...48M} algorithms. They are applied to the SPHERE case as described in \citet{2014A&A...572A..85Z} and in \citet{2015A&A...576A.121M} and they are currently implemented through the SpeCal pipeline \citep{2018A&A...615A..92G}.

\subsection{RV data}
\label{s:rvdata}
Radial velocity data, acquired with the HARPS specrograph, were available from \citet{2020A&A...636A..74T} for HIP\,1481 (24 data points), HIP\,30314 (27 data points),
HIP\,88399 (38 data points) and HIP\,96334 (71 data points).
Unfortunately no RV data, taken with HARPS or any other RV instrument, were available for HIP\,116063. The retrieved RV data were used to obtain mass limits (shown in Fig.~\ref{f:pmaall}) for possible companions at low separations from the host stars using the Exoplanets Detection
Map Calculator \citep[EXO-DMC;][]{2020ascl.soft10008B} and following the method
described in \citet{2021MNRAS.503.1276M}.  

\subsection{PMa data}
\label{pmadata}

All the PMa data used for this work were obtained from the catalogue
by \citet{2022A&A...657A...7K} (see Table~\ref{pmadata}). 
Following their approach we considered as probable companion hosting stars all the targets with a PMa SNR larger than 3. 
Using the method described in \cite{2019A&A...623A..72K} we were able to estimate the mass of the companions compatible with the PMa signal as a function of the separation from the host star (see their Equation~1). The resulting limits are shown in Fig~\ref{f:pmaall} (blue solid lines). It is important to note that the mass is calculated assuming a circular orbit while a statistical distribution for its inclination is taken into account
following the method devised in \citet{2019A&A...623A..72K}. The values obtained should therefore be regarded as a minimum mass for the object causing the PMa signal. Moreover
the possible positions of the companion generating the PMa signal can be
further limited using the position angle (PA) of the velocity anomaly vector $\vec{\Delta v_\mathrm{tan}}$, as shown by \citet{Bonavita2022}. 
We will discuss the resulting constraints in Sec.~\ref{s:conclusion}. 


\section{Results}
\label{s:res}

\subsection{Candidate companions}
\label{s:candidate}

\begin{figure}
\centering
\includegraphics[width=0.9\columnwidth]{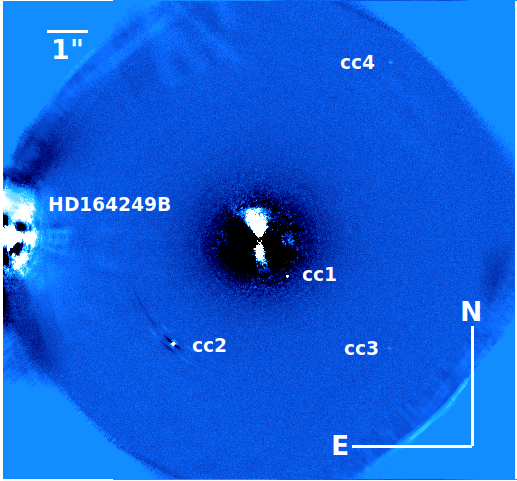}
\caption{Final IRDIS image for HIP\,88399 obtained from data taken on 2018-04-11. The four candidate companions are marked as cc1, cc2, cc3 and cc4, respectively. The known stellar companion HD\,164249\,B is partially visible at the eastern edge of the image.}
\label{f:hip88399}
\end{figure}

\begin{table*}[!htp]
  \caption{Astrometry for the four candidate companions in the IRDIS FOV of HIP\,88399 retrieved from the first and last SPHERE data sets. }\label{t:astrohip88399}
\centering
\begin{tabular}{cccccc}
\hline\hline
cc & Obs. date & Sep. RA (mas) & Sep. Dec (mas) & Total sep. (mas) & PA ($^{\circ}$) \\
\hline
 1 & 2015-05-10 &  -661.50$\pm$6.13 & -1115.98$\pm$6.13 & 1297.30$\pm$8.66 & 210.7$\pm$0.1 \\
 1 & 2019-09-07 &  -649.25$\pm$6.13 &  -758.28$\pm$6.13 &  998.25$\pm$8.66 & 220.6$\pm$0.1 \\
 2 & 2015-05-10 &  2130.28$\pm$6.13 & -2757.48$\pm$6.13 & 3484.50$\pm$8.66 & 142.3$\pm$0.1 \\
 2 & 2019-09-07 &  2142.53$\pm$6.13 & -2418.15$\pm$6.13 & 3230.77$\pm$8.66 & 138.5$\pm$0.1 \\
 3 & 2015-05-10 & -3173.98$\pm$6.13 & -2878.75$\pm$6.13 & 4285.01$\pm$8.66 & 227.8$\pm$0.1 \\
 3 & 2019-09-07 & -3150.70$\pm$6.13 & -2527.18$\pm$6.13 & 4039.00$\pm$8.66 & 231.3$\pm$0.1 \\
 4 & 2015-05-10 & -3197.25$\pm$6.13 &  4120.90$\pm$6.13 & 5125.77$\pm$8.66 & 322.2$\pm$0.1 \\
 4 & 2019-09-07 & -3186.23$\pm$6.13 &  4477.38$\pm$6.13 & 5495.35$\pm$8.66 & 324.6$\pm$0.1 \\
\hline
\end{tabular}
\end{table*}

\begin{figure*}
\centering
  \includegraphics[width=0.8\textwidth]{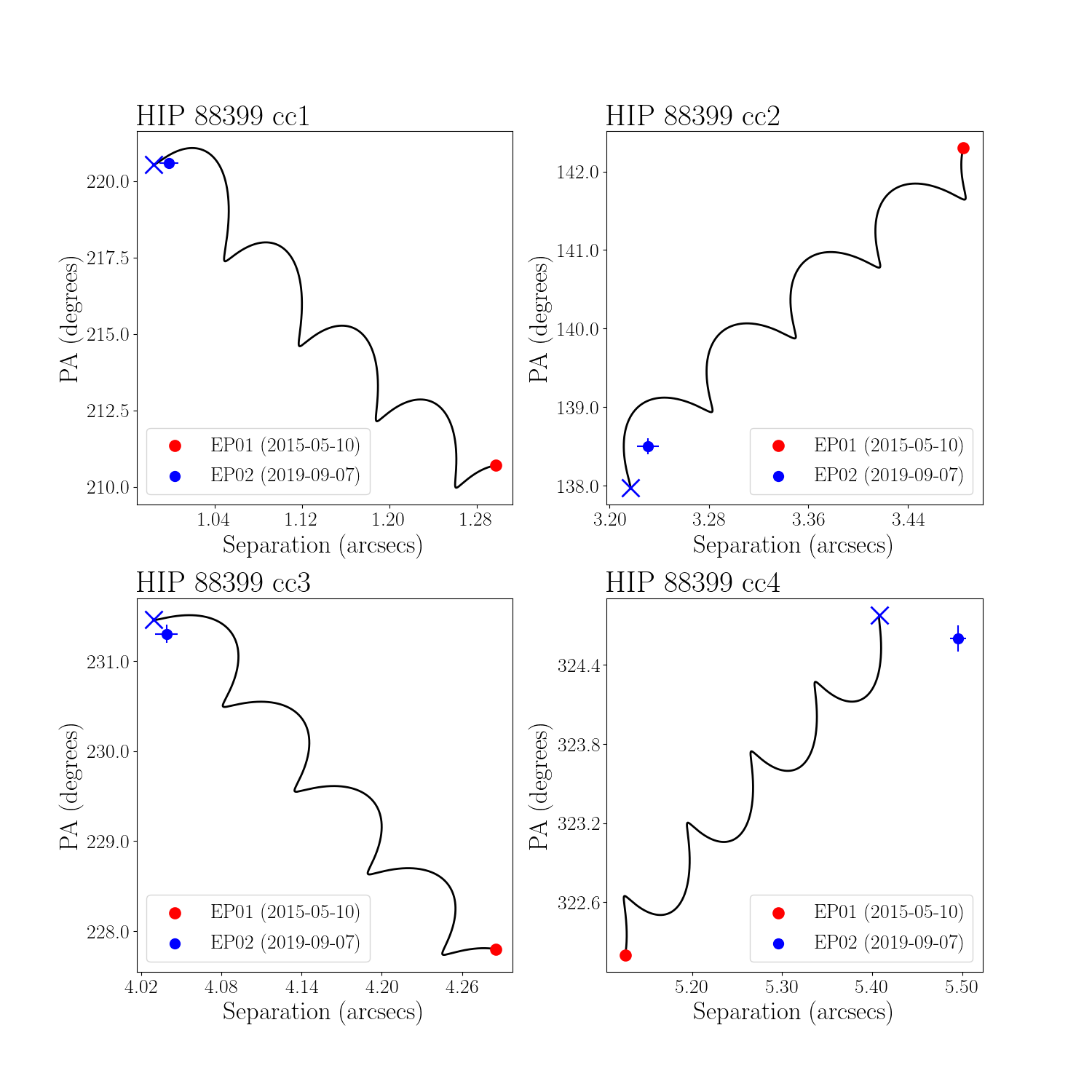}
  \caption{Common proper motion analysis of all the candidate companions identified for HIP~88399. In all panels, the solid black line shows the motion of a background object relative to the target, based on the eDR3 parallax and proper motion of the primary over the same time frame. The filled circles show the measured separation and position angle of the companions at the first (red) and second (blue) epoch. The blue cross indicates the expected position of a background object at the second epoch.}
  \label{f:astrohip88399}
\end{figure*}

\begin{figure}
\centering
\includegraphics[width=0.9\columnwidth]{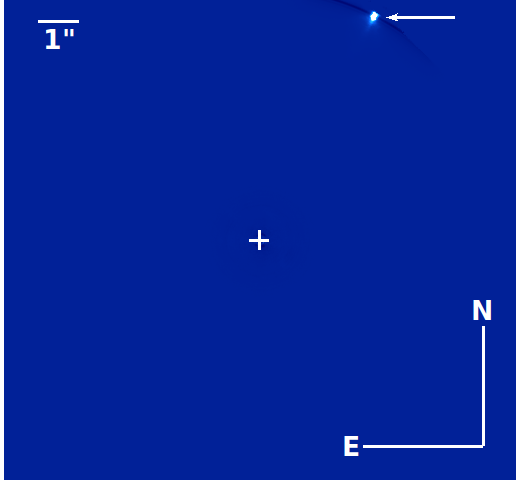}
\caption{Final IRDIS image for HIP\,96334. The candidate companion is
  indicated by the white arrow in the upper part of the image. The position of the
  star behind the coronagraph is indicated by a white cross.}
\label{f:hip96334}
\end{figure}

\begin{table*}[!htp]
  \caption{Astrometry of the candidate companion to HIP\,96334 for the two epochs considered in this work, obtained from NACO and SPHERE observations. }
  \label{t:astrohip96334}
\centering
\begin{tabular}{ccccccc}
\hline\hline
Obs. date  &  Instrument & Pixel scale (mas) & Sep. RA (mas) & Sep. Dec (mas) & Total sep. (mas) & PA ($^{\circ}$) \\
\hline
2007-06-09 & NACO & 27.15 & -2793.74$\pm$13.58 & 3529.50$\pm$13.58 & 4501.37$\pm$19.20 & 321.6$\pm$0.2 \\
2019-05-19 & SPHERE/IRDIS & 12.25 & -2793.00$\pm$6.13 & 5503.93$\pm$6.13 & 6172.04$\pm$8.66 & 333.1$\pm$0.1 \\ 
\hline
\end{tabular}
\end{table*}

\begin{figure}
\centering
\includegraphics[width=0.9\columnwidth]{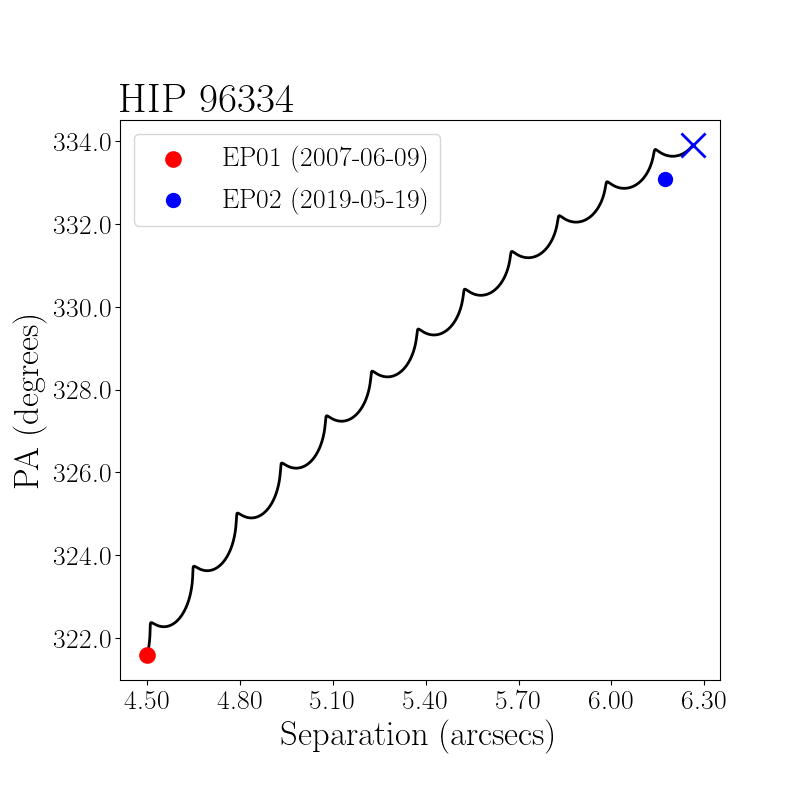}
\caption{Common proper motion analysis for the candidate companion to HIP 96334. The solid black line shows the motion of a background object relative to the target, based on the eDR3 parallax and proper motion of the primary over the same time frame. The filled circles show the measured separation and position angle of the companions at the first (red) and second (blue) epoch. The blue cross indicates the expected position of a background object at the second epoch. }
\label{f:astrohip96334}
\end{figure}
We first inspected the reduced DI data looking for candidate companions that could explain the PMa signal.
In the case of HIP\,1481, HIP\,30314 and HIP\,116063 we did not find any candidate within the IFS or IRDIS FOV. \par
Four candidates were identified in the IRDIS images of HIP\,88399, all at separations larger than 1\as and therefore not included in the IFS images. 
Figure~\ref{f:hip88399} shows the IRDIS image of HIP\,88399, acquired on 2018-04-11 (the night with the best weather conditions), with the candidates highlighted in white. The bright source at the extreme East of the IRDIS image is the known stellar companion: HD\,164249\,B. Given its position, it was however not possible to extract any astrometric value for this star as its PSF was not within the IRDIS FoV.
\par
Table~\ref{t:astrohip88399} includes the values of the {relative astrometry of each candidate, obtained for the first and last epochs available (2015-05-10 and 2019-09-07), which were used to check for common proper motion.  For all the data we adopted a rather conservative error bar of half of the IRDIS pixel scale. In Figure~\ref{f:astrohip88399} we compare these astrometric values
with the relative astrometric position of a background object that is
indicated by a blue cross. This position can be easily calculated starting
from the known proper motion and the parallax of the host star and from
the dates of the observing epochs. A background object with no proper motion
would be expected to be found at the separation and PA indicated by the blue
cross at the epoch of the second observation (indicated by the filled blue
circle). Because a background object will display some proper motion,
perfect correspondence is not expected. For the Figure~\ref{f:astrohip88399}
plots, the filled blue circle is very near the expected position for a
background object. Hence, we disprove the assertion that these objects are
gravitationally bound to the host star. In this latter case we would
expect that the position of the filled blue circle would be almost
coincident with the relative position of the first observation (filled red
circle). Indeed, in this latter case the only source of movement should be
due to orbital motion of the companion that, because of the large separation
of such objects, is expected to be very small. Some examples showing these
plots for gravitationally bound objects could be found, e.g., in Figure~4 of
\citet{Bonavita2022}.  We therefore concluded that none of these objects could
be the cause of the measured PMa. \par
One candidate companion was also identified at the edge of the IRDIS FoV for HIP\,96334, as shown in Figure~\ref{f:hip96334}. Although only one SPHERE epoch was available for this target, we were able to perform a common proper motion analysis (the results of which are shown in Figure~\ref{f:astrohip96334}) using NaCo data available in the ESO archive (Program ID: 079.C-0908(A); PI B.~Zuckerman) taken on the 2007-06-09, where the candidate was also visible. The astrometry values for the two epochs is listed in Table~\ref{t:astrohip96334}. Following
  the same approach described above for HIP\,88399 we were able to confirm that the candidate is a background source, a conclusion strengthened by the detection of the companion in eDR3, where it is listed with a parallax which is inconsistent with that of of HIP\,96334. 

\subsection{Contrast limits}
\label{s:contrast}

\begin{figure*}
    \centering
    \includegraphics[width=\textwidth]{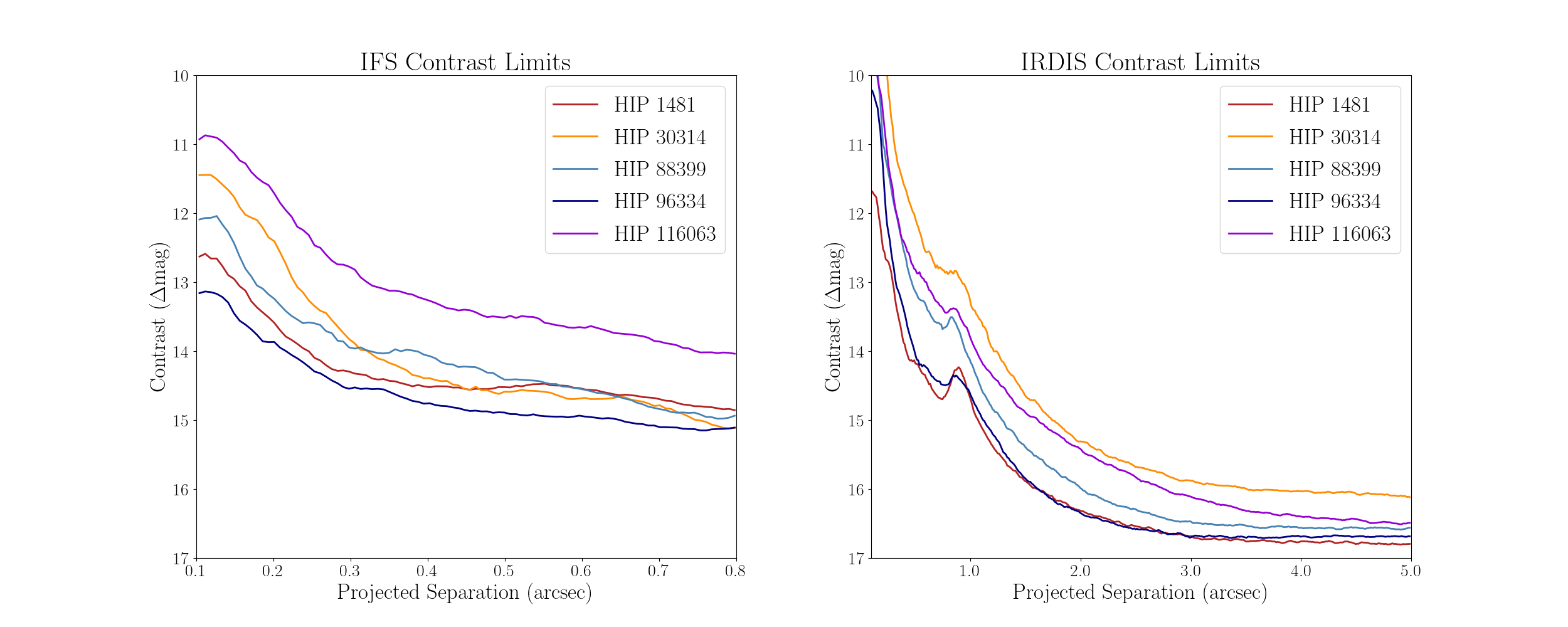}
    \caption{Contrast limits, expressed in magnitude, obtained for all of our targets for IFS (left panel) and IRDIS (right panel) images. In case of multiple epochs, the plotted limit is the one corresponding to the epoch with best weather conditions. Specifically the data taken on the 2016-09-18 were used for HIP\,1481 (red line) and on 2018-04-11 for HIP\,88399 (light blue line). }
    \label{f:contrastall}
\end{figure*}

For each target we calculated the contrast limits for both IFS and IRDIS, using the method described in \citet{2015A&A...576A.121M}. The self-subtraction due to the speckle subtraction method was estimated including simulated objects at different separations from the star and consequently corrected. Finally, the results were corrected
taking into account the effect of the small sample statistics as defined
by \citet{2014ApJ...792...97M}. The resulting contrast limits are shown in Figure~\ref{f:contrastall}.
Note that for HIP\,1481 and HIP\,88399, for which more than one data set was available, we choose to show only the limits obtained using the data taken on the nights with the best weather conditions (2016-09-18 and 2018-04-11, respectively). 

\subsection{Mass limits and comparison with PMa results}
\label{s:masscomp}

\begin{figure*}[htp]
  \centering
  \includegraphics[width=0.9\columnwidth]{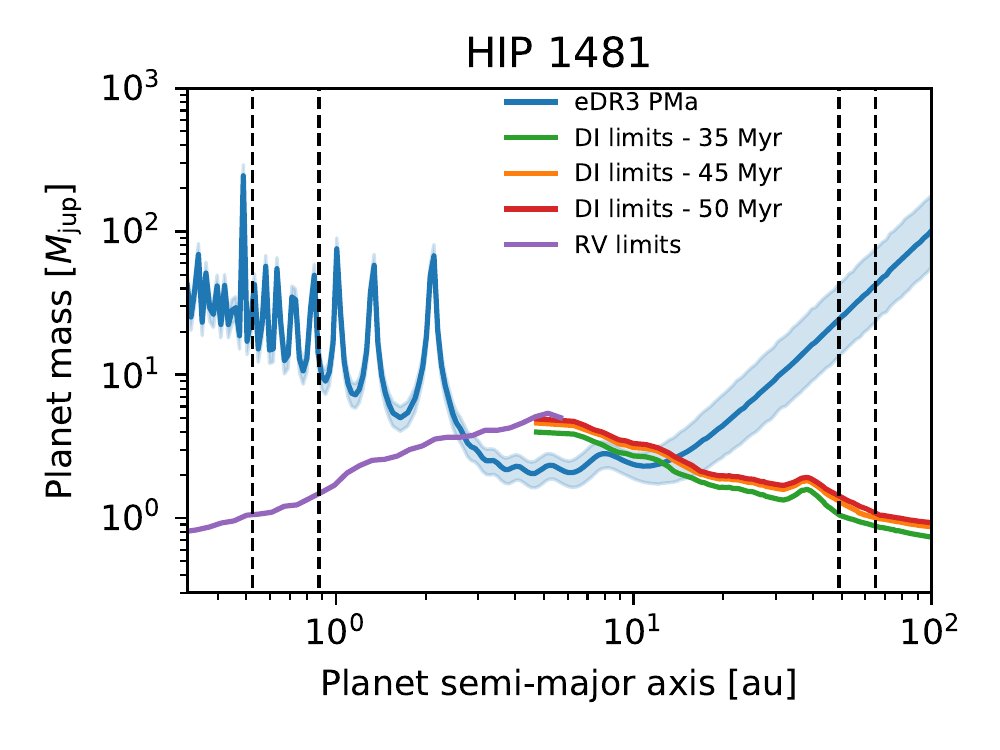}
  \includegraphics[width=0.9\columnwidth]{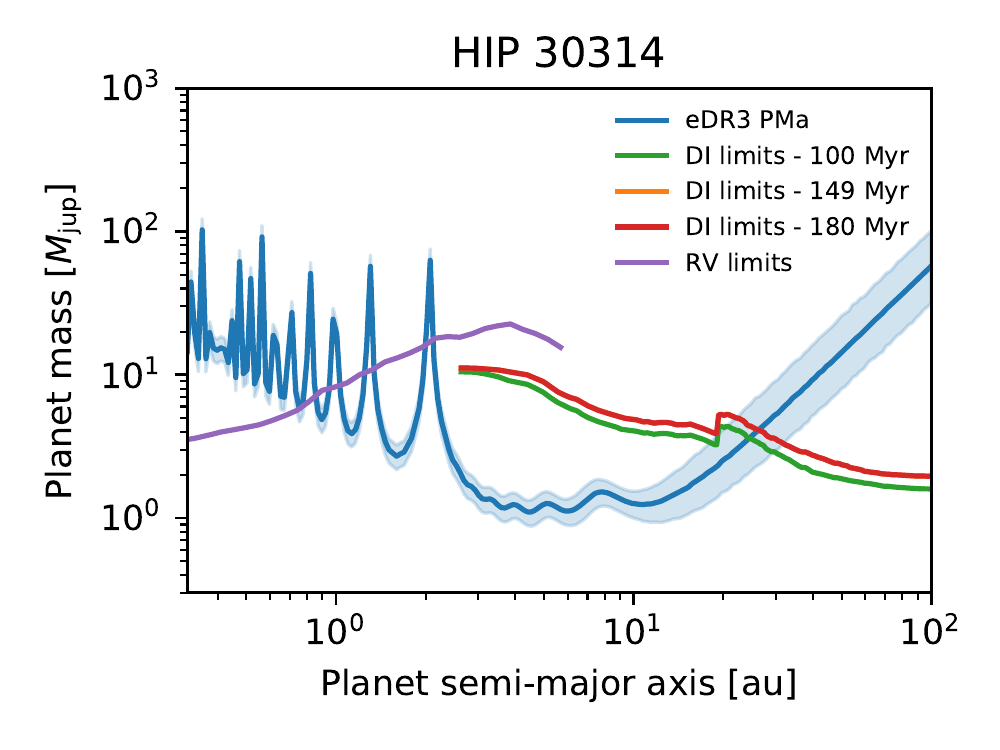}
  \includegraphics[width=0.9\columnwidth]{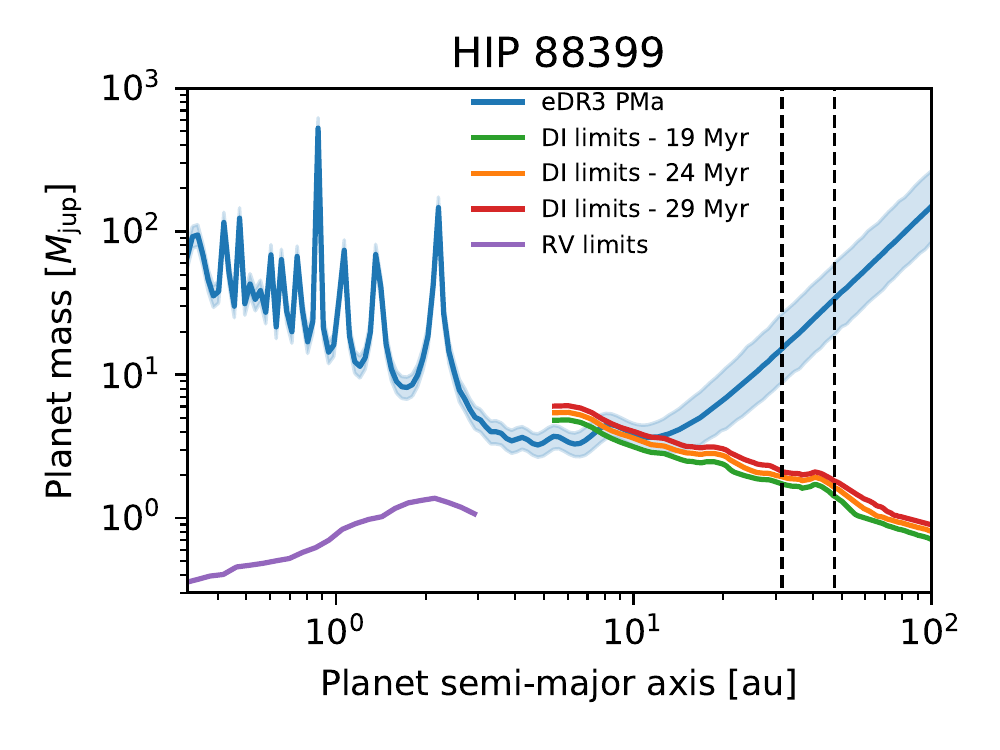}
  \includegraphics[width=0.9\columnwidth]{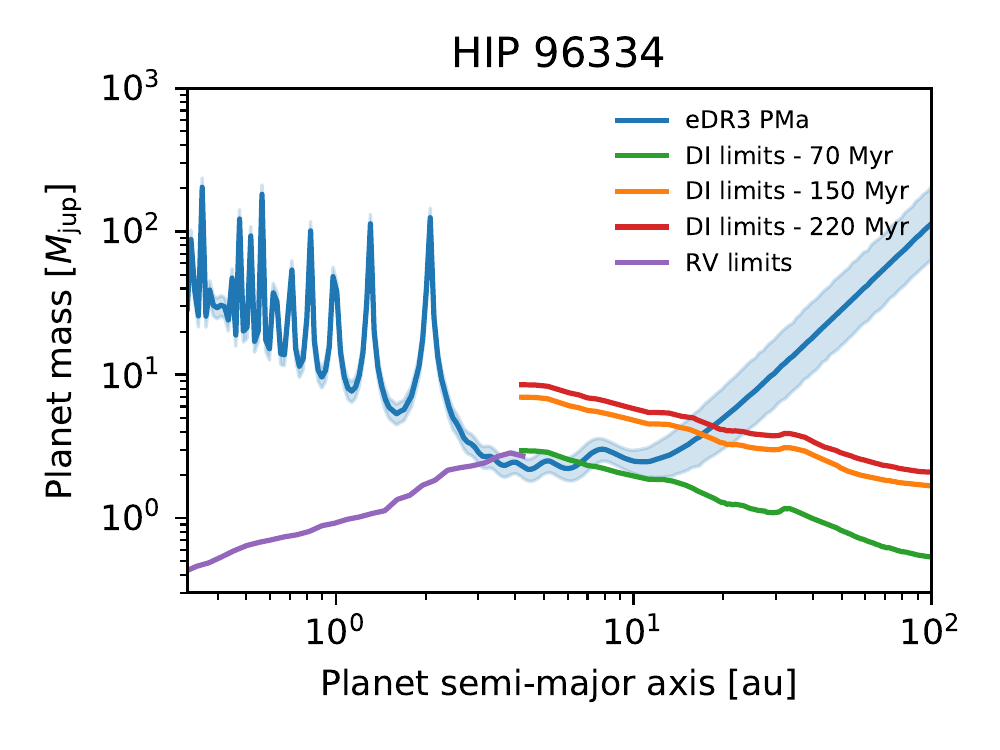}
  \includegraphics[width=0.9\columnwidth]{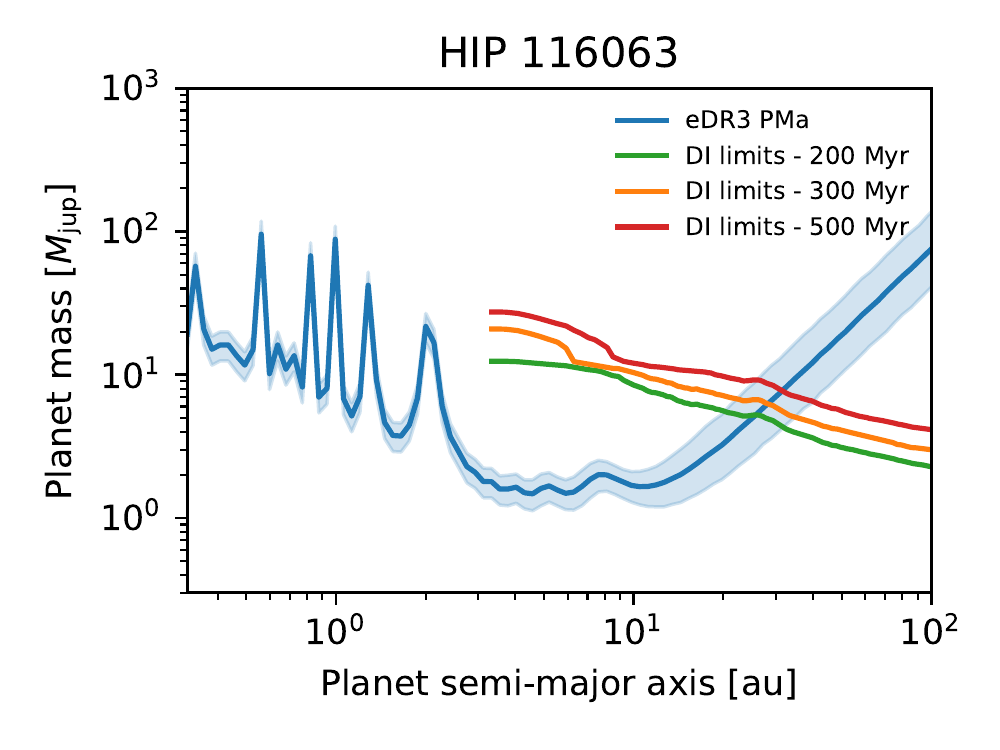}
  \caption{Plots of the mass as a function of the separation from the host
    star of the companion needed to explain the PMa measurement at the Gaia
    eDR3 epoch (blue lines) for HIP\,1481 (top left panel), HIP\,30314 (top
    right panel), HIP\,88399 (center left panel), HIP\,96334 (center right
    panel) and HIP116063 (bottom panel). The blue shaded areas display the
    $1\sigma$ confidence interval. The violet lines represent the mass limits
    from RV data (assuming 95\% confidence level). The DI mass limits assuming
    minimum, expected and maximum ages are shown by the green, orange and red
    lines, respectively. Finally, for HIP\,1481 and HIP\,88399, we also
    included the positions of the belts (two in the cases of HIP\,1481)
    composing the debris disk detected around these stars. They are indicated
    by black dashed lines.}
\label{f:pmaall}
\end{figure*}

The contrast limits shown in Figure~\ref{f:contrastall} were then converted into mass limits using the AMES-COND evolutionary models \citep{2003IAUS..211..325A} and adopting the stellar ages listed in Table~\ref{t:PMa_values}. \par
We then compared these mass limits with those obtained using the RV data (see Section~\ref{s:rvdata}) as well as with the estimates of the mass of the companion responsible for the PMa signal calculated as described in Section~\ref{pmadata}. This allowed us to put further constraints on the mass and the separation of the possible unseen companions. We should at this point stress once more that the mass limits calculated in Section~\ref{pmadata} assume circular orbits and therefore only provide an indication of the order of magnitude for the mass of the companion generating the PMa signal, rather than an exact value. 
The results of the comparison are shown in Figure~\ref{f:pmaall}, where the limits from the PMa are shown as solid blue lines, the RV limits as violet lines and the DI limits are shown with green, orange and red solid lines, depending on the age used for the conversion (minimum, adopted and maximum values, respectively). For HIP\,1481 (top left panel) and HIP\,88399 (center left panel) we also included black dashed lines marking the expected positions of the known debris disks. 


\section{Discussion}
\label{s:dis}

\begin{figure*}
  \centering
  \includegraphics[width=0.8\textwidth]{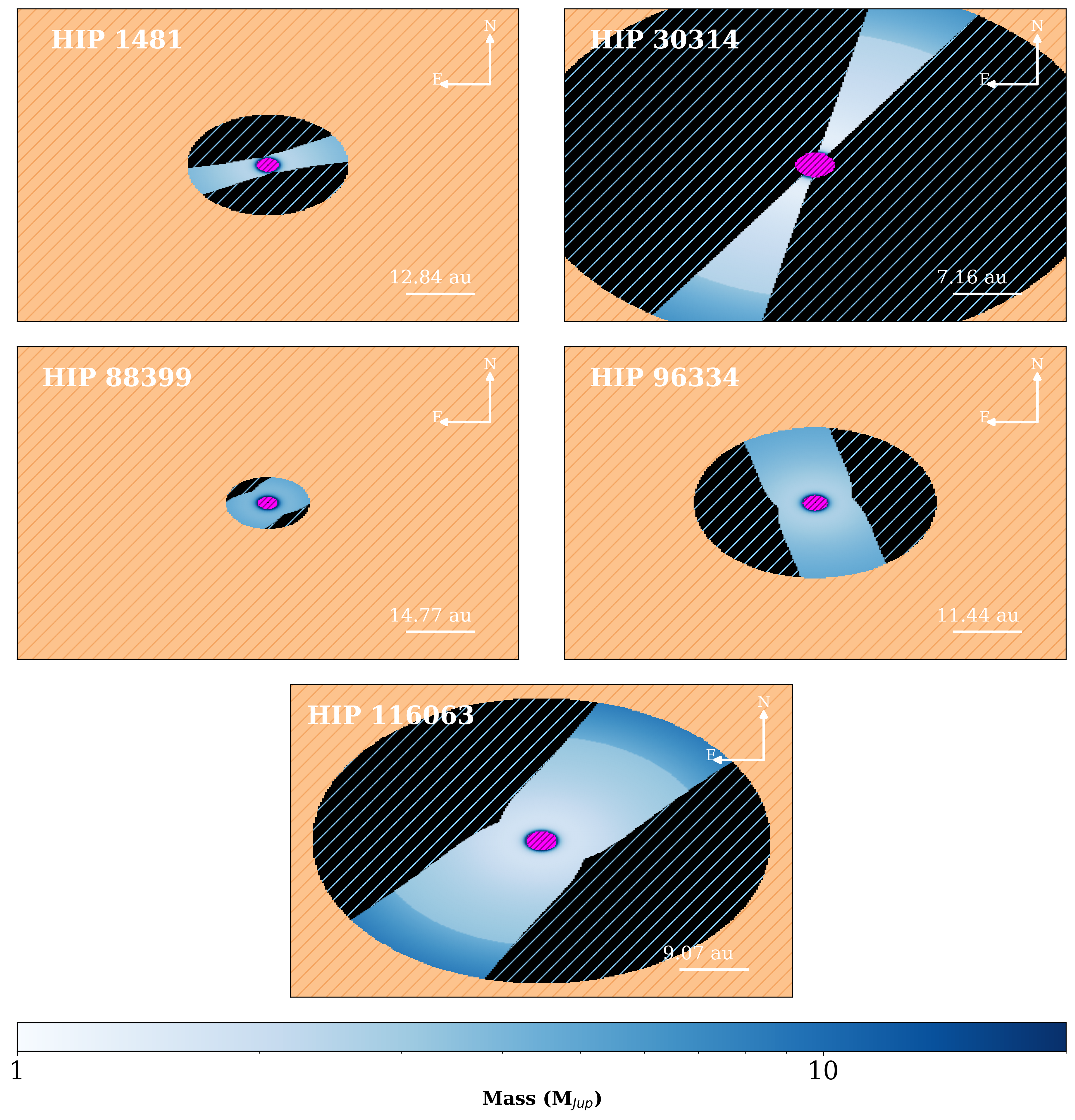}
  \caption{2D maps, obtained with \textsc{FORECAST} \citep[see][for details]{Bonavita2022} showing in blue the sky area compatible with the PMa reported in Table~\ref{t:PMa_values}. The intensity of the blue areas changes according to the dynamical mass (in $M_{Jup}$) responsible for the PMa at a give distance; the same logarithmic scale, shown on the bottom of the figure, was used for all stars. The bar shows the separation corresponding to 0.3\as, expressed in au using the distance of each system. The hashed areas show the regions excluded due to incompatibility with the PMa (black) and using the limits from direct imaging (orange) and radial velocity (purple), respectively (see Fig.~\ref{f:pmaall}).}
\label{f:probmapall}
\end{figure*}

The procedure described in the previous Sections allowed us to determine
different constraints on mass and separations for each of the targets
considered in this work. \par
The most promising results have been obtained for HIP\,1481. As can be seen in
the upper left panel of Figure~\ref{f:pmaall} the DI mass limits exclude the
presence of the companion responsible for the PMa signal at separations larger
than $\sim$15~au while at short separation the RV mass limits exclude the
presence of this companion at separations less than 2-3~au. The possible values of the mass of this object range between 2-5~\MJup with the higher masses possible only at
the lower allowed separations. 
The probability of a companion at separations smaller than $\sim$1~au and at
separations of 40-60~au is further diminished by the presence of the inner
and the outer belt of the debris disk found around this star. \par
The constraints that we can obtain for the case of HIP\,30314 are not as strong (see the upper right panel of Figure~\ref{f:pmaall}). 
Due to the poor weather conditions during the SPHERE observation (see Table~\ref{t:obs}) and the higher age for this system, the DI limits are worse than those achieved for HIP\,1481, especially at short separations (see red and orange solid lines in Fig.~\ref{f:contrastall}), and can only exclude companions at separations larger than $\sim$30~au. 
The limits produced using the RV are also higher than in the case of HIP\,1481, thus providing much less stringent constraints at separations smaller than 1~au. 
The possible masse values range between
$\sim$1~\MJup and 5~\MJup at separation larger than 2~au while they could be
as large as 10~\MJup at separation lower than 2~au. \par
The limits on the separation for HIP\,88399 from the DI data allow to
exclude the presence of the companion causing the PMa signal at separations
larger than 7-9~au (depending on the chosen value of the age). On the other hand, the mass limits
from the RV data are very low and they seem to exclude separations lower than
$\sim$3~au for the companion generating the PMa signal. The mass of the possible
companion ranges between 3 and 5~\MJup with possible
higher masses (up to 8-9~\MJup) at the lower end of the separations range.
As discussed in Section~\ref{s:sample}, this star has a companion with an estimated
mass of 0.54~\MSun (corresponding to $\sim$566~\MJup) and a separation of
$\sim$323~au. 
Looking at the PMa limit in Fig.~\ref{f:pmaall} (blue curve) we can see that at the separation of this companion, the mass requested to explain the PMa signal is of 1547.4~\MJup with a lower limit of 856.3~\MJup, thus well above the estimated mass for the companion. This is true also considering higher masses found in literature
like, e.g, the value of 0.6~\MSun corresponding to $\sim$628.6~\MJup
\citep{2021A&A...651A..70D}. Moreover, the separation used here
for this companion is the projected one. The real separation of this object
could then be even larger, making the difference between its mass and the one
required to explain the PMa signal more evident\footnote{Once again, we note that the
determination of the mass obtained in this way is valid under the
assumption of circular orbits. Eccentric orbits
could broaden the mass distribution at the separation of the companion.}.
As further confirmation, we note also that the PA of the companion ($\sim$90$^{\circ}$) is also not compatible with that derived by \cite{2022A&A...657A...7K} from the PMa ($130.94^{\circ}\pm10.17^{\circ}$), with a difference larger than 3$\sigma$ between the two values.
While from each single value listed above it is difficult to draw a conclusion
because of their large uncertainties, the combination of all
these indications strongly hints toward the exclusion of the known companion
as responsible for the PMa signal. In any case further data, both from astrometry
and DI, is still needed to further clarify the situation. \par
The large errors on the age for HIP\,96334 lead to very different mass limits from the DI images, and this has quite a large impact on the constraints we could put on the nature of the PMa companion. In fact, when considering an age of 70~Myr
(green solid line in the center right panel of Figure~\ref{f:pmaall}), DI data
exclude companions at separations larger than $\sim$7~au. Instead, when considering the much higher ages of 150 and 200~Myr (orange and
red lines in the center right panel of Figure~\ref{f:pmaall}), we could only exclude possible companions at separations larger than $\sim$20~au. On the other hand, mass
limits from the RV allow to exclude any possible companions at separations
lower than $\sim$3~au. From the mass point of view, an age of 70~Myr would allow
for companion of 3-4~\MJup while the higher ages would allow masses
as high as 5-7~\MJup. \par
The lack of RV data limited our ability to put constraints on companions within 1~au from HIP\,116063, where we cannot exclude masses as large as $\sim$100~\MJup. At larger
separations, the bad quality of the DI data and the relatively high age of the
system only allowed us to exclude companions at separation larger than 30~au with
masses that can be as high as 10~\MJup.

\subsection{FORECAST maps}
\label{s:probmap}
The limits shown in Fig.~\ref{f:pmaall} were derived using the absolute value of the PMa reported in Table~\ref{t:PMa_values}. But as mentioned in Section~\ref{pmadata}, further constraints can be put taking advantage of the vectorial nature of the PMa, and in particular of the information on its position angle. 
We therefore used the FORECAST tool \citep[Finely Optimised REtrieval of Companions of Accelerating STars, see][for details and other uses]{Bonavita2022} to isolate the region on the plane of the sky where the companion compatible with the PMa should lie, based on the PA values reported in Table~\ref{t:PMa_values}.

For each target, \textsc{FORECAST} evaluates the position angle of each pixel in the IFS image with respect to the center and then compares it with the PMa position angle (PApma) retrieved from the catalogue by \cite{2022A&A...657A...7K}. 
The optimal region (shown in blue in Fig.~\ref{f:probmapall}) is then identified by the X and Y positions on the image where the position angle is within one sigma from the PApma value, plus or minus an additional quantity which takes into account the possible orbital motion of a companion at that separation between the SPHERE observation and the Gaia observations. 

Finally, it also associates with each point of the resulting 2D map a value of the companion mass based on the PMa absolute value, again using the approach from \citet{2019A&A...623A..72K}, similarly to what was done to obtain the PMa limits shown in Fig.~\ref{f:pmaall}. 
 
Using this information we could further limit the possible positions of the companion on the IFS field of view at the epoch of the observation. Once again, since no information is available about eccentricity and inclination of the orbits, this method only provides a rough indication of the position compatible with circular and edge-on orbits. 
The FORECAST maps can also be used as \emph{finding charts}, to identify the area where to search for possible companions. Figure~\ref{f:probmapall} displays the FORECAST maps obtained for all our targets, considering the IFS field of view. For each panel, the distance bar corresponds to a separation of 0.3\as, which we then converted in $au$ using the distance of the targets. To take into account the constraints derived from Figure~\ref{f:pmaall}, we highlighted the regions excluded thanks to the RV (purple) and SPHERE limits (orange). 
Combined with Figure~\ref{f:pmaall}, Figure~\ref{f:probmapall} therefore provides further visual confirmation that while for HIP\,1481 and HIP\,88399 our limits are relatively tight, for HIP\,30314 and HIP\,116063 the allowed positions area is much larger. \par
The HIP\,1481 and HIP\,88399 systems presents similarities with our solar
system as both of them have a solar type central star with a debris belt
at few tens of au (similar to the Kuiper belt) and a very probable Jupiter-mass
planet at a separation comparable to that of Jupiter. For these
reasons they could be regarded as young analogues of the solar system making
their study even more interesting.

\subsection{Probability of future detection with DI observations}

The SPHERE observations for HIP\,88399 and HIP\,96334 were taken in good
observing conditions so that it is difficult that new observations with
such instruments could allow getting substantial improvement in the
contrast (and in the mass limits). This suggests a low probability of
  direct companion detection using current high-contrast imagers.
The observations for HIP\,1481 were instead taken in intermediate
condition (in particular strong wind, see Table~\ref{t:obs}). Better
  observing conditions could improve the contrast and the mass limits for this
  star helping in detecting the companion or in further constraining its
mass and separation. On the other hand, longer observations should not be
useful to improve the contrast limits as at large angular separations from
the passage at the meridian the rotation of the FOV is low making high-contrast
method like ADI less effective.
Finally, the observations for HIP\,30314 and HIP\,116063 were taken in
bad weather conditions limiting the contrast reached for these targets.
This leaves some room for improvements in constraining the positions and the
mass of the companions with future observations with SPHERE. In any case,
the relatively high ages (larger than 100~Myr) for both of them makes very
challenging the direct detection of the companions especially if they reside
at separations less than 10~au.\\
However, these stars should be ideal targets for future instrumentation at
ELT. We note indeed that the contrasts obtained from simulations of the
performance of ELT first generation instruments like MICADO
\citep[e.g.,][]{2018arXiv180401371P} would allow the detection for the
proposed companions for each separation and mass as constrained by our
analysis. This type of analysis demonstrates then to be a very powerful
tool to select interesting targets for observations with ELT instruments.


\section{Conclusions}
\label{s:conclusion}

This paper presents a detailed analysis based on the combination of DI, RV and astrometric data for five stellar systems showing significant acceleration (PMa) signals suggesting the presence of planetary companions within few au. 
We were able to put constraints on the possible separation and mass of unseen companions on circular orbits for each considered system. \par
Such constraints were particularly strong for HIP\,1481, for which we were able to limit the possible companion separations to 2-15~au and masses of the order of few \MJup. 
Also interesting are the cases for HIP\,88399 and HIP\,96334 for which similar constraints can be obtained despite of the limits imposed by the strong uncertainty on the age. 
For HIP\,88399, although more data are needed for final confirmation, our analysis seems to exclude the known wide companion HD\,164249\,B as the reason of the PMa signal. \par
Finally, looser constraints were possible for HIP\,30314 and HIP\,116063. The bad weather condition in which the DI observations were taken in both cases, and the lack of RV data in the second case, strongly limited our
capacity to put constraints both on mass and separation. 
Despite such difficulties, these targets still proved to be interesting, with a good probability of
the presence of a companion at separations between few au and up to 20-30~au.
\par

While the characteristics of the companions compatible with the PMa strongly limit the feasibility of a detection with current current high-contrast instruments (e.g., SPHERE and GPI), they represent ideal targets for observations with ELT-class telescopes, as their proposed masses and separations perfectly fit with the estimated performance of the instruments planned for these facilities.


\begin{acknowledgements}
This work has made use of the SPHERE Data Center, jointly operated by
OSUG/IPAG (Grenoble), PYTHEAS/LAM/CeSAM (Marseille), OCA/Lagrange (Nice) and
Observatoire de Paris/LESIA (Paris). \par
This work has made use of data from the European Space Agency (ESA) mission
{\it Gaia} (\url{https://www.cosmos.esa.int/gaia}), processed by
the {\it Gaia} Data Processing and Analysis Consortium (DPAC,
\url{https://www.cosmos.esa.int/web/gaia/dpac/consortium}). Funding for
the DPAC has been provided by national institutions, in particular the
institutions participating in the {\it Gaia} Multilateral Agreement. \par
This research has made use of the SIMBAD database, operated at CDS,
Strasbourg, France. \par
This work has been supported by the PRIN-INAF 2019 "Planetary systems at young ages (PLATEA)" and ASI-INAF agreement n.2018-16-HH.0. A.Z. acknowledges support from the CONICYT + PAI/
Convocatoria nacional subvenci\'on a la instalaci\'on en la academia,
convocatoria 2017 + Folio PAI77170087. \par
SPHERE is an instrument designed and built by a consortium consisting
of IPAG (Grenoble, France), MPIA (Heidelberg, Germany), LAM (Marseille,
France), LESIA (Paris, France), Laboratoire Lagrange (Nice, France),
INAF-Osservatorio di Padova (Italy), Observatoire de Gen\`eve (Switzerland),
ETH Zurich (Switzerland), NOVA (Netherlands), ONERA (France) and ASTRON
(Netherlands), in collaboration with ESO. SPHERE was funded by ESO, with
additional contributions from CNRS (France), MPIA (Germany), INAF (Italy),
FINES (Switzerland) and NOVA (Netherlands). SPHERE also received funding
from the European Commission Sixth and Seventh Framework Programmes as
part of the Optical Infrared Coordination Network for Astronomy (OPTICON)
under grant number RII3-Ct-2004-001566 for FP6 (2004-2008), grant number
226604 for FP7 (2009-2012) and grant number 312430 for FP7 (2013-2016).\par
For the purpose of open access, the authors have applied a Creative Commons Attribution (CC BY) licence to any Author Accepted Manuscript version arising from this submission. 

\end{acknowledgements}

\bibliographystyle{aa}
\bibliography{propermotion5}

\end{document}